\documentclass[sigconf]{acmart}
\AtBeginDocument{%
  }

\acmYear{2026}\copyrightyear{2026}
\acmConference[CAIS '26]{Proceedings of the 1st ACM Conference on Agentic and AI Systems}{May 26, 2026}{San Jose, CA, USA}
\acmBooktitle{Proceedings of the 1st ACM Conference on Agentic and AI Systems (CAIS '26), May 26, 2026, San Jose, CA, USA}
\acmDOI{10.1145/3786335.3813223}
\acmISBN{979-8-4007-2415-2/26/05}

\begin{document}

\title[\System{}: Dynamic Autonomy for Coding Agents Under Local Oversight]{\System{}: Dynamic Autonomy for Coding Agents Under\\ Local Oversight}

\author{Tanjal Shukla}
\affiliation{%
  \institution{University of Washington}
  \city{Seattle}
  \state{WA}
  \country{USA}
}
\email{tshukl@uw.edu}

\author{K. J. Kevin Feng}
\affiliation{%
  \institution{University of Washington}
  \city{Seattle}
  \state{WA}
  \country{USA}
}
\email{kjfeng@uw.edu}

\author{Leijie Wang}
\affiliation{%
  \institution{University of Washington}
  \city{Seattle}
  \state{WA}
  \country{USA}
}
\email{leijiew@cs.washington.edu}

\author{Mohammad Rostami}\authornote{This work is independent of the position at Amazon.}
\affiliation{%
  \institution{Amazon GenAI Innovation Center}
  \city{Seattle}
  \state{WA}
  \country{USA}
}
\email{rostamii@amazon.com}

\author{Amy X. Zhang}
\affiliation{%
  \institution{University of Washington}
  \city{Seattle}
  \state{WA}
  \country{USA}
}
\email{axz@cs.uw.edu}

\renewcommand{\shortauthors}{Shukla et al.}

\begin{abstract}
Despite coding agents' advances in handling increasingly complex tasks, their continued tendency to introduce unintended edits, subtle bugs, and scope drift that slip past code review means developers must still decide how much autonomy to grant them.
However, existing approaches for setting an agent's level of autonomy, such as static permission settings or instruction files, cannot account for how developers' preferences for agent autonomy can shift across tasks and over time.
We conducted a formative survey with 21 software engineers who use coding agents and found that they experience frustration with calibrating autonomy and have evolving preferences for level of oversight. Building on these insights, we present \System{}, a CLI coding agent that dynamically adjusts its autonomy level based on developer-agent interactions across sessions. Rather than operating on a global, fixed autonomy configuration, \System{} learns an evolving set of behavioral guidelines from developer decisions and feedback, reducing friction on work for which the agent has earned trust, while tightening oversight when the agent operates outside familiar territory. 
\System{} demonstrates the potential of a new paradigm where 
agents intelligently adapt their level of autonomy based on user trust 
through active, longitudinal collaboration.

\end{abstract}



  \begin{teaserfigure}
  \centering
  \includegraphics[width=1\textwidth]{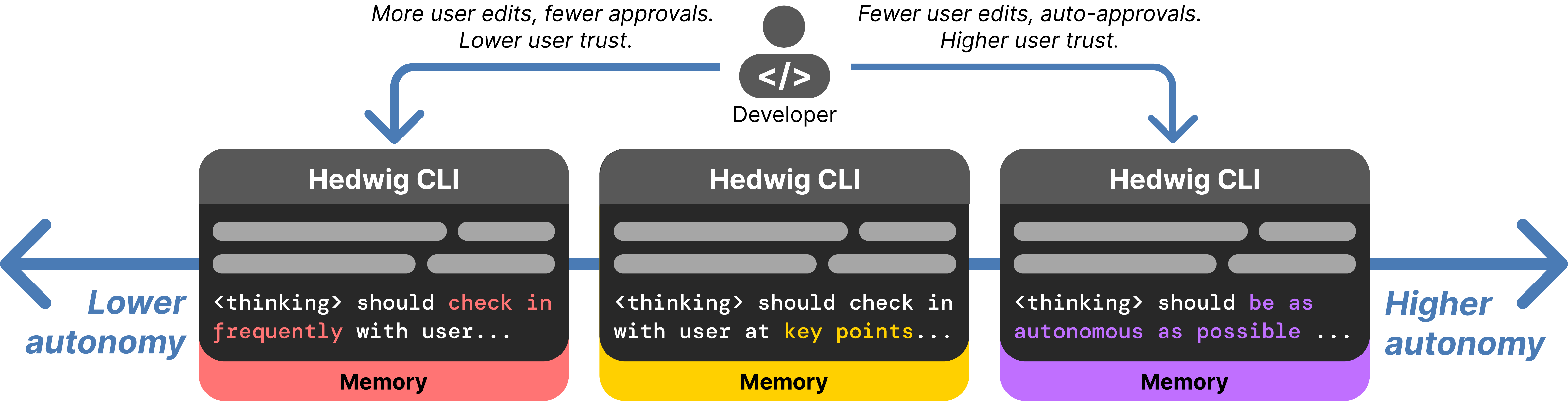}
  \caption{\System{} is a CLI coding agent that dynamically calibrates its level of autonomy in response to user interactions and feedback. \System{} records and reasons over user  interactions (e.g., code edits, plan corrections, rejected/approved commands) over time to maintain an evolving set of behavioral policies. These policies are used to govern the agent’s actions and frequency of check-ins with the developer.}
  \label{fig:teaser-abstract}
\end{teaserfigure}



\newcommand{\System}{\textsc{Hedwig}}
\maketitle


\section{Introduction}                                                                 
Coding agents are increasingly integrated into everyday programming practices. Modern coding agents inspect repositories, edit multiple files, run commands, and debug errors---capabilities that require deep access to developers' working environments and can create risks of unintended edits, subtle bugs, or security issues. Field observations of professional developers confirm that effective use of agents is highly controlled in practice: developers retain responsibility for design decisions, verification, and code review rather than delegating end-to-end autonomy~\cite{huang2025dontvibe}. Consequently, collaborating with a coding agent is not only about judging the correctness of its outputs, but also effectively managing the agent through strategic calibration of its level of autonomy~\cite{feng2025levels}.

Yet what level of autonomy is desired can vary across developers, tasks, timelines, and projects. Some developers want to review every edit, while others prefer to step in only at key milestones or when evaluating the final artifact. Although today's coding agents are shaped by large-scale post-training and alignment pipelines meant to capture generally useful developer preferences, these defaults often fail to match the autonomy expectations of a particular user in a specific project context~\cite{feng2024cocoa, Peng2025MoraePP, Mozannar2025MagenticUITH}. As a result, the goal of human-AI collaboration should not be to maximize automation, but to maintain meaningful human oversight---calibrated to each developer's risk tolerance, task stakes, and evolving trust in the agent ~\cite{bansal2021appropriate}. This calibration requires understanding how oversight behavior actually varies across users and tasks~\cite{huq2026cowcorpus}.

Researchers and practitioners have explored several approaches for calibrating agent autonomy. Coding agents are often designed to ask for clarification or permission before taking risky actions, but the right frequency and timing of these check-ins is non-trivial: empirical work shows that deferring all confirmation to the end is suboptimal---81\% of users preferred intermediate checkpoints in a recent 48-participant study~\cite{zhou2026checkin}---yet frequent interruptions can disrupt workflow and become burdensome~\cite{grunde2026overseeing, Mozannar2025MagenticUITH}. Developers can also specify behavioral preferences through instruction files (e.g., \texttt{AGENTS.md}), profile settings, or stored memories that guide future interactions. However, these approaches often assume that users' preferred level of autonomy is stable and can be articulated in advance. In practice, autonomy preferences \textit{shift over time} as users gain or lose trust in the agent, move across tasks with different stakes, or work under different constraints. To achieve this, high-quality feedback from human developers can be a more reliable signal than upfront preferences alone~\cite{liang2026pahf}. This points to a need for coding agents that can intelligently learn from interaction and continually adapt their level of autonomy.

We present \System{}, a CLI coding agent that \textit{dynamically adapts its autonomy level} based on developer interactions. We designed \System{} based on insights from a formative survey with 21 professional programmers with extensive experience with using coding agents. Rather than fixing its autonomy at a particular configuration, \System{} reasons about and adapts its autonomy level over time based on how the developer interacts with the agent in practice.

\System{} has two main components. First, it records developer-agent interactions and  categorizes them into two governance tiers: \emph{hard constraints} enforced directly by the CLI, and \emph{behavioral guidance} retrieved when relevant to shape how the model approaches the current task. 
Behavioral guidance is stored as compact feedback snippets and logic notes from related past work---capturing developer corrections, architectural decisions, and prior reasoning---and retrieved into future prompts when relevant to the current task, allowing the model to reuse prior decisions and corrections. Second, when the coding agent proposes reads, plans, check-ins, or code updates, \System{} audits those proposals against active hard constraints, retrieved behavioral context, the current task and specification, and interaction history (e.g., approvals, denials, and corrections). Over time, this history trains an online policy that adjusts when similar future actions require a check-in. Beyond instruction-file approaches, hard constraints are enforced at the CLI layer independently of the model's reasoning, and the end-of-session summary separates policy-triggered from model-triggered check-ins---making each oversight decision legible to the developer.

\section{Motivations and Design Goals}
 
\subsection{Motivations from a Developer Survey}
\label{s:formative}
We conducted a formative survey with 21 experienced users of coding agents, which motivated the need for selective, legible oversight. Our sample mostly consisted of experienced engineers: 20 of 21 respondents reported at least 8 years of programming experience, and 20 of 21 used coding agents at least weekly. Their leading concerns were subtle bugs that pass review (13/21), erosion of their own understanding of the codebase (12/21), security vulnerabilities (10/21), and outdated or incorrect context (9/21). At the same time, respondents did not want constant interruption: 11 of 21 preferred check-ins at key milestones, and another 2 of 21 preferred check-ins only when something risky or unexpected occurred. These results point to the same design requirement: coding agents need oversight that is selective, legible, and adjustable rather than uniformly strict or uniformly permissive~\cite{grunde2026overseeing,huq2026cowcorpus,liang2026pahf,wang2025humansmissing,bansal2021appropriate}. This study was reviewed and approved by the University of Washington IRB.

\subsection{Design Goals}

We derive three design goals using insights from results of our developer survey, which guide our design and development of \System{}.

\textbf{DG1: Support adaptive autonomy through interactions between developers and coding agents.} Recent coding agents can persist project instructions (e.g, AGENTS.md) and accumulate preference notes (e.g., Claude Code Auto Memory), but these mechanisms primarily reload generic prior guidance as context for the model. They do not directly use developers' observed interactions as a signal for runtime oversight. In practice, however, developers often discover their preferred level of oversight only after working with an agent over time~\cite{grunde2026overseeing, feng2024cocoa}. Their preferences may change as they build trust, encounter mistakes, or move across tasks with different stakes. This makes it difficult for developers to fully specify their desired level of autonomy in advance. Instead, the agent should adapt over time based on how developers actually respond to its behavior in practice.

\textbf{DG2: Strategically calibrate and communicate oversight.} Current coding agents typically support oversight in two broad ways: either the runtime enforces interruptions (e.g., whenever the agent needs to run a terminal command), or the model is prompted to ask for help (e.g., asking clarification questions). As our survey suggests, when the agent ``checks in'' with the user is often not tailored to individual preferences and usage patterns, and the basis for such interruptions can be opaque. An agent should decide more strategically when to involve the user and when to proceed autonomously, using interaction history to reduce unnecessary interruptions while still surfacing important decisions. It should also communicate why a check-in is needed, or why less oversight is required in familiar situations.

\textbf{DG3: Enable the agent to flexibly move between levels of autonomy.} From our survey, we learned that developers do not want one uniform level of control over coding agents. The same programmer may want strict oversight for some files or actions, but much lighter oversight for others. Their preferences may also differ by task, risk level, or stage of work. In addition, some preferences are hard boundaries that should be deterministically enforced, while others are softer behavioral guidelines about when and how the agent should involve the developer. Existing agentic systems may offer prompt instructions, memory, permission settings, and hooks, but developers must often decide manually which mechanism to edit and how each rule should be enforced. A coding agent with adaptive autonomy should make this distinction explicit and enforceable at runtime rather than solely relying on the model's interpretation.

\section{The \System{} CLI Agent}

\subsection{Technical Components}

\begin{figure*}[t]
  \centering
  \includegraphics[width=1\textwidth]{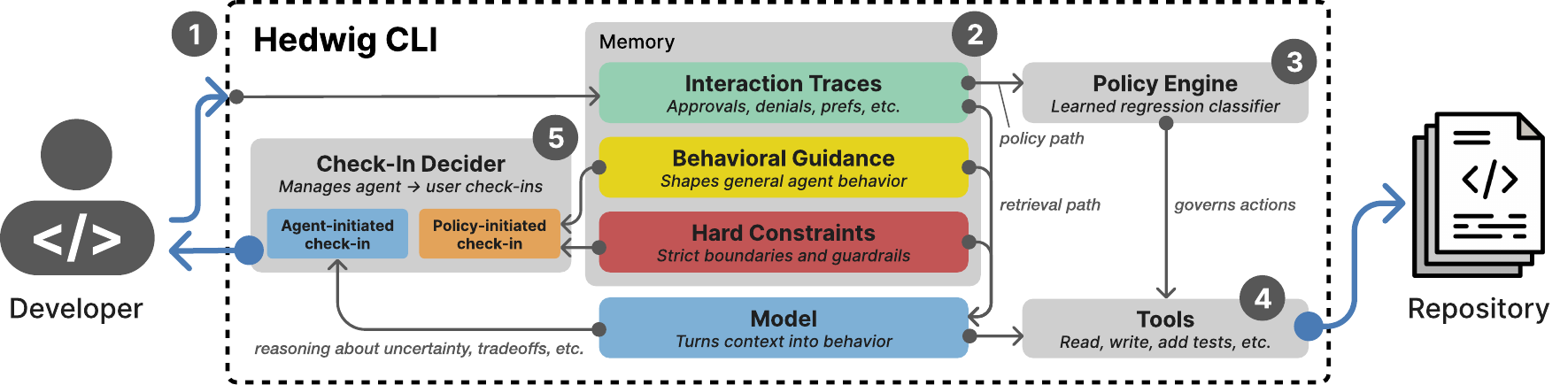}
  \caption{Architecture of \System{}. (1) A developer's interactions with the agent are stored in the agent's memory (2). Also in the memory module are \textit{hard constraints} and softer \textit{behavioral guidance}. The model can retrieve from memory via the \textit{retrieval path} to inform its behavior. Separately, the \textit{policy path} trains a policy engine (3), a regression classifier. The model can ask to use tools (4), and tool use is also governed by the policy engine. Finally, \System{} uses the behavioral guidance and hard constraints to partially inform when and how to check in with the user (5). The model can also reason about this independently.}
  \label{fig:architecture}
\end{figure*}

\System{} is a local Python CLI that supports dynamic autonomy in collaboration between developers and coding agents. It observes interaction traces over time and maintains an evolving set of autonomy policies based on that history. These policies regulate the agent's actions and surface check-ins to the developer at appropriate moments, along with brief explanations. \System{} is powered by a cloud-hosted LLM (specifically, Claude Sonnet 4.6) and integrated into VS Code. We provide \System{}'s source code for local usage and testing.\footnote{Source code: \url{https://github.com/tanjalshukla/HedwigCLI}}

\textbf{Categories of autonomy policies.} Mirroring how developers think about autonomy in their collaboration with coding agents, \System{} distinguishes two kinds of autonomy policies (\textbf{DG2}). \emph{Hard constraints} capture strict boundaries on what the coding agent should or should not do, such as avoiding edits to a protected path or always involving the developer before modifying it. \emph{Behavioral guidance} captures softer preferences for how the agent should collaborate, such as when to be cautious, when to ask for input, or how much initiative to take. Autonomy preferences are bidirectional: while trust broadens as consistent approvals accumulate, developers can selectively revoke a specific granted preference---for a particular file area or change category---without discarding all other stored state (\textbf{DG1}). This ensures oversight can tighten, not only loosen, as a developer's trust posture changes.

\textbf{Inferring autonomy policies from recorded traces.} \System{} listens to a range of interaction traces between the developer and the agent, such as approvals or denials of proposed actions, how carefully the developer reviews changes, corrections to agent outputs, revisions to agent plans, verification outcomes, and explicit developer guidance. Each interaction drives two parallel feedback paths that remain architecturally separate. In the \emph{policy path}, trace history trains an online logistic regression classifier that directly governs future check-in behavior. The classifier is initialized from engineering priors via a synthetic warm-start and updated via stochastic gradient descent after each developer decision, operating over a 13-dimensional feature vector capturing properties of each proposed action: diff size, blast radius, change pattern risk, prior approval and denial counts, security sensitivity, verification failure history, and model confidence, among others. Coefficients are persisted per repository so personalization accumulates across sessions. The classifier operates entirely on the CLI side; the model is never shown numeric scores or learned weights. In the \emph{retrieval path}, developer corrections and explicit instructions are stored as short text snippets and retrieved into future prompts based on relevance to the current task, allowing the agent to draw on prior decisions qualitatively without access to the underlying policy state. In other words, the same action may require approval early on, but later proceed automatically once the agent has accumulated evidence that similar actions were safe and acceptable in that part of the repository. For example, 3 denials on an API file shift \texttt{change\_pattern\_risk} and \texttt{prior\_denials} toward higher scrutiny, while 5 approvals on a utility file lower the check-in bar for that path across future sessions.

\textbf{Applying behavioral guidance.} As one form of autonomy policy, behavioral guidance helps steer how the coding agent approaches future tasks. At the start of each run, \System{} retrieves the stored corrections, behavioral guidelines, and logic notes from related past work that are most relevant to the current task, and includes only those in the prompt given to the coding agent (\textbf{DG1}). This qualitative retrieval is distinct from the learned policy: it shapes how the agent reasons, while the policy engine independently governs what the agent is permitted to do.

\textbf{Enforcing hard constraints.} To enforce hard constraints at runtime, \System{} grants authority at the level of individual actions (e.g., reads and writes) as opposed to a global permission setting~\cite{lampson1971protection}. Every proposed action is evaluated through a three-tier approval cascade: hard constraints are checked first, followed by the learned policy engine. The resulting score determines one of three outcomes: actions above the proceed threshold are approved silently and appear only in the session summary; actions in an intermediate band are approved but surfaced with a compact summary line for post-run inspection; only actions below the flag threshold trigger a live check-in. Stored autonomy preferences and their resulting thresholds are directly inspectable via \texttt{hw observe preferences}. To ensure a quality outcome before completion, any approved change is also checked against the project's test suite as configured by the developer. In addition, \System{} tracks the coding agent's current workflow phase---such as research, planning, implementation, or review---and only allows actions appropriate to that stage.

\textbf{Surfacing check-ins with clear explanations.} \System{} surfaces check-ins at moments when developer input is needed to refine or apply autonomy policies (\textbf{DG2}). Some check-ins are triggered by the system based on hard constraints, milestone triggers, or risk threshold conditions. Others are triggered by the coding agent itself upon detecting uncertainty, conflict with prior guidance, plan deviation, or a meaningful design tradeoff. These two sources operate in parallel but remain categorically distinct; the end-of-session summary reports policy-initiated and agent-initiated check-ins separately, making the two-tier architecture directly legible to developers after every run. During each check-in, \System{} also provides a brief explanation of why the check-in was triggered, including how it relates to prior interaction traces when relevant.

\textbf{Personalizing interactions with developers.}
The mechanisms described above allow \System{} to personalize interactions with developers who have different levels of tolerance for risk when programming with coding agents. At a high level, coefficient updates from online training of \System{}'s regression classifier captures diverging behaviors and preferences from different developer personas, enabling the agent to tailor its actions effectively; we illustrate this concretely in Appendix~\ref{a:personalization}.

\section{Evaluation: Learning Check-in Behavior from Developer History}
\label{sec:eval}

We evaluate whether Hedwig's learned policy adapts its check-in frequency to a developer's historical approval patterns, and compare against Claude Code (CC) with Auto Memory. Both systems completed two identical backend tasks---(T1) adding a \texttt{GET /tasks/summary} endpoint and (T2) extending it with an optional \texttt{priority} filter---yielding 11 write operations per run. For Hedwig, we seeded the trust database with 20 synthetic decisions from a \emph{cautious} or \emph{permissive} developer persona and replayed them 3$\times$ (60 effective decisions) to reach a stable classifier state. Seeding additionally infers autonomy preferences from approval rate: personas with $\geq 90\%$ approval trigger \texttt{prefer\_fewer\_checkins} and \texttt{skip\_low\_risk\_plan\_checkpoint}. CC's Auto Memory stores preferences as prose but does not produce a quantitative check-in threshold, so we encoded the same personas as rules in \texttt{CLAUDE.md}; CC-B is the no-rules control. To ground each system's per-operation decisions we also score them against an LLM judge (Claude Opus~4.7) given the persona rules and operation context; the judge labels 4 operations across T1+T2 as requiring a check-in under the cautious persona (two new public additions, two signature changes) and 0 under the permissive persona.

\begin{table}[h]
\small
\centering 
\setlength{\tabcolsep}{4pt}
\caption{Check-ins and alignment with an LLM-judge oracle (Claude Opus~4.7), summed across both tasks (11 ops total). Oracle requires 4 check-ins under the cautious persona, 0 under permissive. Recall = required caught / required; Precision = required caught / caught. Both are undefined (---) when the denominator is zero. Bolded values highlight each axis's best result.}
\label{tab:eval}
\begin{tabular}{llrrrr}
\toprule
Run & Persona & Chk & Ratio & Recall & Precision \\
\midrule
H-C  & cautious   & 6 & 0.55 & \textbf{1.00} & 0.67 \\
H-P  & permissive & 4 & 0.36 & --- & 0.00 \\
\midrule
CC-C & cautious   & 2 & 0.18 & 0.50 & \textbf{1.00} \\
CC-P & permissive & 0 & 0.00 & --- & --- \\
CC-B & ---        & 0 & 0.00 & 0.00 & --- \\
\bottomrule
\end{tabular}
\end{table}

The two systems occupy different points on a safety-efficiency tradeoff. \textbf{On the cautious persona, Hedwig's recall is 1.00 vs.\ CC-C's 0.50}: CC-C silently passes through two signature-change operations that its own \texttt{CLAUDE.md} explicitly instructed it to review, while Hedwig catches all four judge-required check-ins at the cost of two false positives. In a governance context a missed check-in (un-reviewed API change ships) is costlier than an extra one (two seconds of developer time), so Hedwig's recall advantage is the safety-dominant metric. \textbf{On the permissive persona, CC wins on precision}: CC-P and CC-B produce zero false positives against a zero-check-in oracle, while Hedwig incurs four. Hedwig's floor here is by design---first-time file reads and structural API-surface modifications retain governance even for trusted developers---but it can be a real efficiency cost. Full setup, per-finding analysis, and limitations are in Appendix~\ref{a:eval_full}.

\subsection{System Walkthrough}
We illustrate \System{} through a two-session scenario on a small Python task API with route handlers, a service layer, tests, and a short task specification file. The specification acts as a task contract, marking API-facing behavior that should not change without approval while leaving implementation strategy open. To make cross-session personalization visible within the live demo slot, the demo repo is pre-seeded with 14 synthetic developer decisions so the learned policy is active from the first \texttt{hw run}. The same scenario is shown in the accompanying demo video.\footnote{Demo video: \url{https://www.youtube.com/watch?v=1OGPF1iVeJs}}

The developer begins by adding two natural-language rules using \texttt{hw rules add}. The first---``never write to files under \texttt{config/prod/}''---is compiled into a hard constraint enforced directly by the CLI. The second---``for routine backend changes, reuse existing validation helpers and avoid creating new files unless clearly necessary''---is classified as behavioral guidance and persisted for future retrieval. \System{} displays its interpretation of each rule for confirmation before persisting it, making the governance classification explicit and transparent.

\textbf{Session 1.} The agent requests to read three files. The developer approves with remember (\texttt{r}), establishing reusable read access for the session. The model raises a planning check-in before any edits are made, surfacing a handler signature tradeoff: query-based (\texttt{summary\_handler(query: dict)}, consistent with existing patterns) versus parameter-less (\texttt{summary\_handler()}, returning all summaries directly---simpler but breaking the existing convention). This is an agent-initiated check-in the model paused on its own to surface a design tradeoff, rather than a CLI policy rule firing. The developer selects the query-based option and provides brief architectural guidance: preserve the nested response structure, prefer fewer interruptions for low-risk internal changes, and continue checking in only for API, signature, schema, or security changes. \System{} produces a plan grounded in the task specification, touching \texttt{task\_api/api.py} and \texttt{task\_api/service.py}, presents it for approval, applies the patch, and runs the configured verification command. Verification passes. The end-of-session summary reports: \textit{Check-ins: 3 policy-initiated, 1 agent-initiated.}

Between sessions, \texttt{hw observe preferences} surfaces the stored autonomy preferences and effective scoring bands: actions scoring above 0.90 are approved silently, actions between 0.20 and 0.90 are approved but flagged for post-run inspection, and actions below 0.20 trigger a live check-in. \texttt{hw observe weights} shows learned coefficients against warm-start priors; with the pre-seeded history incorporated, the two largest shifts are \texttt{model\_confidence\_avg} ($+0.51$) and \texttt{change\_pattern\_risk} ($+0.13$), reflecting that this developer's history penalizes API-pattern changes and rewards high model confidence.

\textbf{Session 2.} A related follow-up task extends the summary endpoint with an optional priority filter. The read stage proceeds immediately without prompting: \System{} displays \textit{``read: reused prior read access on 2/2 files; prior approvals 6; prior denials 0 -- guidance: `Use the nested object response. Preserve the existing response...'\thinspace''} The retrieved guidance from session 1 has been injected into the model's prompt context, narrowing uncertainty for the follow-up task and suppressing a repeat of the architectural check-in. Friction is reduced \emph{selectively}: the plan checkpoint still fires, because balanced mode preserves a policy-initiated gate for multi-file plans. The developer approves the plan. The apply stage requires one manual approval since remembered write access was not yet established on these files. The session ends with \textit{Check-ins: 3 policy-initiated.}---one fewer agent-initiated check-in than Session 1, consistent with the resolved design uncertainty.

\textbf{Observability and trust control.} Running \texttt{hw observe report} after both sessions shows the full governance trace: 1 agent-initiated check-in with a 100\% approval rate, 10 policy check-ins, 7 deliberate approvals, and 4/4 verification passes. The initiator breakdown---separating policy-driven interruptions (CLI rules and scoring thresholds) from agent-driven ones (model uncertainty or design tradeoffs)---is a first-class output rather than something buried in trace exports. This separation lets the developer see whether interruptions came from too-strict system rules or genuine model uncertainty. Running \texttt{hw observe preferences-revoke --topic api} removes the API topic from trusted scope without resetting accumulated states, firing a check-in on all future API changes regardless of prior approval history. This allows oversight to tighten and loosen accordingly.

\System{} is built as a proof-of-concept demo for dynamic autonomy in agentic systems. Its strongest contribution is architectural: it externalizes oversight into a continually-updated, inspectable policy while leveraging model-side reasoning as a useful but non-authoritative source of how and when to collaborate with the user. This creates a reusable framework for studying the longitudinal evolution of autonomy in coding agents.

\System{} has several limitations. The evaluation is based on synthetic interaction traces derived from our formative survey; no formal user study has been conducted. The approval policy is initialized from engineering priors and refined via online SGD; convergence rates and persona-specific learning trajectories (including when and how quickly a permissive profile's check-in rate should approach zero on routine operations) remain open questions for future study. The system does not yet implement reversibility of a code change as a risk signal, incorporate a richer preference taxonomy, or enable sub-agent delegation. Its spec-driven development support is simplistic: the model is grounded in a spec digest and reports requirement coverage, but no structured pipeline is provided.
 
The present artifact is a working prototype and study platform, not a production-grade personalization method. Integrating Zhou et al.'s CDCR-aware scheduling~\cite{zhou2026checkin} as a timing layer above \System{}'s learned policy is a natural direction. The most immediate next step is a controlled lab study evaluating not only task completion but interruption burden, steerability, verification effort, and calibrated reliance during realistic workflows~\cite{wang2025humansmissing}.

\section{Conclusion}

We introduce \System{}, a CLI coding agent that intelligently calibrates its level of autonomy based on developer interactions. \System{} records and reasons over user interaction traces (e.g., edits to proposed, rejected/approved commands, corrections to its plans) over time to maintain an evolving set of policies that govern agent behavior. Specifically, these policies are used to decide how and the frequency of which the agent seeks user assistance and feedback. Future work includes conducting user studies with developers to validate our design and crafting more sophisticated governance mechanism for autonomy beyond a text-based policy.





\section{Acknowledgements}
We would like to thank our anonymous reviewers for their insightful feedback. We also would like to express our heartfelt thanks to all the participants who dedicated their time and effort to participate in our study. Research reported in this publication was supported by an Amazon Research Award, Fall 2024.

\bibliographystyle{ACM-Reference-Format}
\bibliography{refs}


\appendix
\section*{Appendix}
\section{Synthetic Trace Generation}
\label{a:synthetic}

Each persona was defined as a profile specifying which change patterns, diff sizes, blast radii, and review behaviors trigger approval or denial. Synthetic decisions were generated by instantiating \texttt{PolicyInput} feature vectors matching each profile's behavioral rules and assigning labels (approve/deny) consistent with the persona's documented preferences. Twenty decisions were generated per persona and replayed 3$\times$ during training (60 effective gradient updates), covering a range of scenarios including edge cases where the persona's rules produce non-obvious outcomes (e.g., a large test-generation diff that a Cautious developer approves because change pattern risk is low). A fresh warm-start classifier was then trained on each persona's decision set and coefficient deviation from the warm-start priors was measured.

\subsection{Personalizing Developer Interactions}
\label{a:personalization}

To make the personalization mechanism concrete, we constructed three developer personas spanning the trust spectrum. The exercise illustrates how coefficient updates produce different policies for different developers; it is not a quantitative evaluation of system performance.

\textbf{Personas.} Each persona is a behavioral profile specifying which action features trigger approval or denial: a \emph{Cautious} developer denies API and schema changes and reviews carefully; a \emph{Permissive} developer approves broadly and intervenes only on security-sensitive changes; a \emph{Mixed} developer approves tests and documentation freely but denies large multi-file diffs. These three were chosen to span the trust postures observed in our formative survey---one risk-averse, one risk-tolerant, and one selective.

\textbf{Features and units.} The four features in Table~\ref{tab:weight_shift} are:
\begin{enumerate}
    \item \texttt{change\_pattern\_risk}, a risk score derived from the type of change (API and schema modifications score high, documentation and tests score low).
    \item \texttt{model\_confidence\_avg}, the model's self-reported confidence in the proposed action.
    \item \texttt{is\_security\_sensitive}, a binary flag for actions touching security-related paths.
    \item \texttt{prior\_denials}, tallying previous denials on similar actions.
\end{enumerate}

Each cell in Table~\ref{tab:weight_shift} reports $\Delta = \text{learned} - \text{prior}$, the unitless shift in the classifier's logistic regression coefficient on that feature relative to its warm-start value. Positive $\Delta$ means the feature now pushes more strongly toward auto-approval; negative $\Delta$ pushes toward a check-in.

\begin{table}[h]
\small
\centering
\caption{Policy coefficient deviation ($\Delta$ = learned $-$ prior) after 15 synthetic decisions per persona. Values are unitless log-odds coefficient shifts on standardized features.}
\label{tab:weight_shift}
\begin{tabular}{lccc}
\toprule
\textbf{Feature} & \textbf{Cautious} & \textbf{Permissive} & \textbf{Mixed} \\
\midrule
\texttt{change\_pattern\_risk}   & $+0.24$ & $-0.33$ & $+0.28$ \\
\texttt{model\_confidence\_avg}  & $+0.27$ & $+0.81$ & $+0.49$ \\
\texttt{is\_security\_sensitive} & $-0.06$ & $-0.20$ & $-0.12$ \\
\texttt{prior\_denials}          & $-0.27$ & $-0.83$ & $-0.50$ \\
\bottomrule
\end{tabular}
\end{table}

The same feature (\texttt{change\_pattern\_risk}) shifts $+0.24$ for Cautious and $-0.33$ for Permissive, while \texttt{model\_confidence\_avg} drifts positive across all three personas---the classifier learns developer-specific weights rather than averaging toward a global default. \texttt{is\_security\_sensitive} shows the sharpest negative drift for Permissive ($-0.20$), correctly isolating the single signal that persona acts on.

\section{Full Evaluation Setup, Findings, and Limitations}
\label{a:eval_full}

This appendix expands Section~\ref{sec:eval} with full setup detail, per-finding analysis, and limitations.

\subsection{Setup}

For Hedwig, the two personas are: \emph{cautious} (100\% denial on API/data-model/config changes with slow, deliberate review; 100\% approval on tests/docs/general changes) and \emph{permissive} (19/20 approvals across all change patterns, with a single denial on an extreme multi-file refactor). Each persona's 20-decision set is replayed 3$\times$, producing 60 effective gradient updates on the online logistic-regression classifier. Approval rate in the seed set also drives autonomy-preference inference: the permissive persona's 95\% approval rate triggers \texttt{prefer\_fewer\_checkins} and \texttt{skip\_low\_risk\_plan\_checkpoint}, which relaxes the pre-implementation plan gate for low-risk plans (scope-only gating; hard-risk signals such as security-sensitive paths, low-trust files, and strict mode still block bypass regardless of preferences).

For Claude Code, we tested three conditions: CC-C (\texttt{CLAUDE.md} says ``ask before API-shape changes''), CC-P (\texttt{CLAUDE.md} says ``proceed autonomously on routine backend edits''), and CC-B \\(no \texttt{CLAUDE.md}). We verified that CC behaves the same whether \texttt{CLAUDE.md} sits at project root or \texttt{\textasciitilde/.claude/}, and temporarily renamed the user's pre-existing global configuration out of the load path so it did not confound the trials.

A \emph{check-in} is an agent-initiated, unprompted approval request before a write. CC's harness-level permission prompts (``Allow this Edit?'') are excluded, as they fire on every tool call regardless of configuration and are therefore not persona-sensitive by construction.

\textbf{LLM judge.} To ground per-operation decisions against an independent labeler, we scored each of the 11 operations against Claude Opus~4.7 given the persona rules and operation context (file, change type, new-function/endpoint status, signature modification). The judge labels 4 operations as requiring a check-in under the cautious persona (adding \texttt{summarize\_tasks()} and \texttt{summary\_handler()} in T1; modifying both signatures in T2) and 0 under the permissive persona. We report \emph{recall} as the fraction of judge-required check-ins the system issued; recall is undefined for permissive runs because the oracle expects zero check-ins, so for those rows we report false-positive counts in the prose. All counts in Table~\ref{tab:eval} are sums across T1 and T2 within a single run per condition; we did not aggregate across repeated runs and so report no median, mode, or variance.

\subsection{Claude Code Persona Prompts}
\label{a:cc_prompts}

The following \texttt{CLAUDE.md} files were used as hand-authored configuration for the CC-C and CC-P conditions. Both files were placed at \texttt{demo\_task\_api/CLAUDE.md} during their respective trials; we verified behavior was invariant to project-root vs.\ \texttt{\textasciitilde/.claude/} placement.

\textbf{Cautious persona (CC-C):}

\begin{quote}
\small\ttfamily
\# Developer Preferences\\[0.3em]
I am a cautious developer. I want to review all API changes before they are applied.\\[0.3em]
- Always ask for my approval before modifying any public handler signatures in \texttt{task\_api/api.py}.\\
- Always ask for my approval before changing any function signatures in \texttt{task\_api/service.py}.\\
- Always ask for my approval before adding new endpoints or new public functions.\\
- Always ask for my approval before changing validation logic or error handling.\\
- For low-risk changes like adding comments, renaming internal variables, or formatting, you may proceed without asking.\\
- Do not create new files without my approval.\\
- Prefer checking in over proceeding autonomously when in doubt.
\end{quote}

\textbf{Permissive persona (CC-P):}

\begin{quote}
\small\ttfamily
\# Developer Preferences\\[0.3em]
I am a permissive developer. I trust you to make routine changes autonomously.\\[0.3em]
- You may modify handler signatures, add endpoints, and update service functions without asking.\\
- You may add, remove, or change validation logic without asking.\\
- You may create new files if clearly necessary without asking.\\
- Only ask for my approval if a change touches security-sensitive logic, authentication, or data persistence in a way that could cause data loss.\\
- Proceed autonomously for all routine backend changes including API additions, refactoring, and test updates.\\
- Prefer proceeding over checking in when in doubt.
\end{quote}

CC-B used no \texttt{CLAUDE.md} file.

\subsection{Findings}

\textbf{(1) Hedwig adapts to developer history without configuration.} The check-in ratio differs across personas (0.55 cautious vs.\ 0.36 permissive, a 19-point gap) purely from a different trust database and its approval-rate-derived autonomy preferences.

\textbf{(2) CC under-checks relative to what its own instructions specify.} Against the LLM judge, Hedwig catches all 4 cautious-required check-ins (recall 1.00) with 2 false positives---over-cautious, consistent with a cautious developer's preference. Claude Code catches only 2 of 4 (recall 0.50), silently skipping the two signature-change operations in T2 despite \texttt{CLAUDE.md} explicitly instructing review on signature changes. CC-B catches 0 of 4 (recall 0.00), confirming that without a hand-authored rule there is no cautious signal at all. The aggregate check-in counts therefore \emph{understate} the safety gap: Hedwig's 0.55 rate is over-cautious-by-design, while CC-C's 0.18 rate is under-cautious-by-mechanism (50\% miss rate on the operations the persona actually wanted reviewed). We did not observe CC's Auto Memory translating observed developer behavior into check-in thresholds. CC additionally supports \texttt{PreToolUse} hooks that can enforce deterministic gating, but these must still be hand-coded per-project and do not learn from approval history.

\textbf{(3) Permissive efficiency tradeoff.} On the permissive persona the oracle expects 0 check-ins. CC-P and CC-B produce 0 (perfect against the permissive oracle) while Hedwig-Permissive incurs 4 false positives. The design tradeoff is explicit: Hedwig treats ``permissive approval history'' as a signal to bypass the plan gate on low-risk plans, not as a signal to disable all governance on structurally risky operations (API-surface modifications, even in trusted contexts). CC's floor is lower at the cost of silently passing through operations the cautious oracle flagged.

\subsection{Limitations}

This evaluation is illustrative, not statistically conclusive. (i)~Hedwig begins with 60 seeded decisions plus inferred autonomy preferences; CC runs cold---we claim Hedwig \emph{has} a per-project learning mechanism, not a head-to-head win from identical starting states. (ii)~Single run across two tasks; no variance estimate across repeats. (iii)~Hedwig check-ins are policy-initiated (learned risk threshold and plan-gate bypass eligibility), CC's are model-initiated via prompting---equivalent in user effect, not mechanism. (iv)~CC's harness prompts are excluded because they are not content-sensitive; counting them would equalize CC conditions and obscure the configuration-driven signal. (v)~CC reads both \texttt{CLAUDE.md} and \\ \texttt{AGENTS.md}; we observed format-dependent behavior on identical content and report \texttt{CLAUDE.md} results to avoid conflating this with the main effect. (vi)~The LLM judge (Claude Opus~4.7) was given each persona's rules and per-operation context; judge labels depend on how the persona rules are interpreted, and a reviewer using a different judge model or rubric could derive different labels. (vii)~Hedwig's autonomy-preference inference uses a fixed approval-rate threshold ($\geq 90\%$); a sliding-window or recency-weighted inference is natural future work and would better reflect drifting developer preferences. (viii)~All CC trials used Sonnet~4.6 on Claude Code~v2.1.119. (ix)~Our tasks are routine backend edits; behavior on higher-risk operations (authentication, data migrations) may differ.
\clearpage

\end{document}